\def\mearth{{\rm\,M_{Earth}}}
\def\rearth{{\rm\,R_{Earth}}}
\def\msun{{\rm\,M_{Sun}}}

\def\lsun{{\rm\,L_{Sun}}}
\def\mjup{{\rm\,M_{Jup}}}
\def\gsim{~\rlap{$>$}{\lower 1.0ex\hbox{$\sim$}}}
\def\lsim{~\rlap{$<$}{\lower 1.0ex\hbox{$\sim$}}}

\def\wpmsq{W m$^{-2}$}

\def\eg{{\it e.g.\ }}

\def\ie{{\it i.e.\ }}
\def\cf{{\it c.f.\ }}

\documentclass[12pt, preprint]{aastex}
\usepackage{graphicx}
\usepackage{appendix,amsmath}
\begin{document}

\title{Habitable Planets Around White and Brown Dwarfs: The Perils of
a Cooling Primary}
\author{Rory Barnes\altaffilmark{1,2,3},Ren{\' e} Heller\altaffilmark{4}}

\altaffiltext{1}{Astronomy Department, University of Washington, Box 951580, Seattle, WA 98195}
\altaffiltext{2}{NASA Astrobiology Institute -- Virtual Planetary
Laboratory Lead Team, USA}
\altaffiltext{3}{E-mail: rory@astro.washington.edu}
\altaffiltext{4}{Leibniz Institute for Astrophysics Potsdam (AIP), An der Sternwarte 16, 14482 Potsdam, Germany}

\begin{abstract}
White and brown dwarfs are astrophysical objects that are bright
enough to support an insolation habitable zone 
(IHZ). Unlike hydrogen-burning stars, they cool and become less
luminous with time, and hence their IHZ moves in with time. The
inner edge of the IHZ is defined as the orbital radius at which a
planet may enter a moist or runaway greenhouse, phenomena that can
remove a planet's surface water forever. Thus, as the IHZ moves in,
planets that enter it may no longer have any water, and are
still uninhabitable. Additionally, the close proximity of the IHZ to
the primary leads to concern that tidal heating may also be strong
enough to trigger a runaway greenhouse, even for orbital
eccentricities as small as $10^{-6}$. Water loss occurs due to
photolyzation by UV photons in the planetary stratosphere, followed by
hydrogen escape. Young white dwarfs emit a large amount of these
photons as their surface temperatures are over $10^4$~K. The situation
is less clear for brown dwarfs, as observational data do not constrain
their early activity and UV emission very well.  Nonetheless, both types
of planets are at risk of never achieving habitable conditions, but
planets orbiting white dwarfs may be less likely to sustain life than
those orbiting brown dwarfs. We consider the future habitability of
the planet candidates KOI 55.01 and 55.02 in these terms and find they
are unlikely to become habitable.
\end{abstract}

\section{Introduction \label{sec:intro}}

The search for extrasolar life begins with the search for liquid
water. All life on Earth requires liquid water and hence it is
reasonable to surmise that life beyond the Earth is also sustained by
it. Most campaigns to search for life have focused on the habitable
zone (HZ) of hydrogen burning ``main sequence'' (MS) stars that are
similar to the Sun. However recent studies have begun to explore the
possibility of habitable planets around brown dwarfs (BD)
\citep{AndreeshchevScalo04,Bolmont11} and white dwarfs (WD)
\citep{Monteiro10,Agol11,Fossati12}.  Furthermore, as these objects are relatively small and dim, transits are more readily detectable than for MS stars \citep{Blake08,Faedi09,DiStefano10,Agol11}. In this investigation we consider the
evolution of the HZs of these objects in the context of tidal heating
and the evolution of the primary's luminosity and spectral energy
distribution.

WDs are the remnants of stars that have burned all the hydrogen in
their cores into helium. As the hydrogen burning ends, the core begins
to fuse the helium atoms and becomes hotter. This extra heat blows the
envelope away, leaving just the core. Most WDs are about the size of
the Earth, but have a mass of $\sim 0.6~\msun$ and a luminosity of
$\sim 10^{-3}~\lsun$ \citep{Bergeron01}. However, a range exists, and
the luminosity actually evolves over several orders of magnitude.

BDs are objects that are not massive enough to produce the central pressures
necessary to fuse hydrogen into helium. Nonetheless, BDs
can support an HZ as their slow contraction converts
gravitational energy into heat \citep{Burrows97,Baraffe03}. The
habitability of planets about BDs has received less attention
than habitability of planets orbiting hydrogen-burning stars, as
a) relatively few are known, b) it is unknown if planets can form around
them, and c) they are very faint. However, new surveys, such as that
of the WISE spacecraft, have found hundreds of these objects
\citep{Mainzer11,Kirkpatrick11}, opening the possibility of detecting
planets in orbit around BDs.

Although no terrestrial planets are currently known to orbit either a WD or a
BD, there is no reason to believe they do not exist. Giant planets
have been detected, such as a $\sim
5~\mjup$~planetary companion $\sim 55$~AU from a nearby BD \citep{Chauvin05}. Subsequent observations have confirmed this object is a
companion to the BD \citep{Song06}, but its colors and luminosity
still pose a challenge to accurately estimating its mass
\citep[\eg][]{Gizis07,Skemer11}. \cite{Mullally08} used the periodicity in arrival times of pulsations of one white dwarf, GD66, to infer the presence of
a $2.4~\mjup$ object at 4.5~AU. Follow-up observations have only
placed upper limits on the companion's mass, leaving its planetary
status ambiguous \citep{Mullally09,Farihi12}. Additional giant planets
on wide orbits halve also been detected around the evolved (post-MS)
stars NN Serpentis
\citep{Beuermann10,Horner12}, DP Leonis \citep{Beuermann11}, and HW
Virginis \citep{Beuermann12}.

Furthermore, there is ample evidence for protoplanetary disks that
could transform into terrestrial planets. BDs have been observed to
host protoplanetary disks that could form planets
\citep{Jayawardhana03,Apai05}. Some WDs host metal-rich disks
\citep{Gansicke06}, and others appear to have been polluted with metals
or water, possibly from tidally-disrupted planets or asteroids
\citep{Jura03,JuraXu12}. While the post-MS evolution is a challenging
barrier to the survival of close-in planetary companions to WDs
\citep[see \eg][]{Nordhaus10}, some or all of the rocky core may
survive engulfment, or planets may form from the debris of the stellar
envelope. For example, two  terrestrial planet candidates have recently been
discovered orbiting Kepler Object of Interest (KOI) 55, a $\sim
0.5~\msun$ ``extreme horizontal branch'' star that will likely become
a WD \citep{Charpinet11}. Thus, we assume that water-rich terrestrial
planets can exist around these objects.

Previous work on BD and WD HZs
\citep{AndreeshchevScalo04,Monteiro10,Agol11,Bolmont11} determined
orbits for which the radiation flux of the primary is equal to the
flux limits found for MS stars
\citep{Kasting93,Selsis07,Barnes08_hab}. The identification of
this so-called insolation habitable zone (IHZ) is an
important first step in constraining the possibility of habitable
planets orbiting WDs and BDs, but one cannot neglect how a
planet behaves prior to the arrival of the IHZ at its orbit. The inner
edge of the HZ is defined to be the orbits at which a desiccating
greenhouse, either moist, runaway or tidal, is just possible
\citep{Kasting93,Barnes12_TV}. These phenomena may ultimately remove
all surface water and leave an uninhabitable planet behind. Thus,
planets initially interior to the HZ may not actually be habitable
after the primary has cooled and/or tidal heating has
subsided sufficiently for the planet to reside in
it. Note that we will refer to the HZ as resulting from both
irradiation and tidal heating, but to the IHZ as the classic
habitability model of \eg \cite{Kasting93}.

For desiccation to occur, a multi-step process must transpire. First,
the surface flux must reach a critical value $F_{crit}$, which is
typically around
300~\wpmsq~\citep{Kasting93,Abe93,Selsis07,Pierrehumbert10}. At this
level, water may either escape the troposphere, and/or become opaque
to infrared radiation. In both cases, water vapor in the stratosphere
can then be photolyzed by high energy radiation. Then, the freed
hydrogen atoms may escape to space \citep{Watson81}. Without the
hydrogen to bond with oxygen, the planet has no water and is not
habitable. Extrapolating from the Solar System, such planets will be
Venus-like with large CO$_2$-dominated atmospheres. Thus, in order to
become sterile, the planet must also be bathed in high energy
radiation for a long enough time for all the hydrogen to be lost. 
\cite{Barnes12_TV} argued for a desiccation timescale, $t_{des}$ of
$10^8$ years for low mass stars. The situation is more complicated
here as BDs and WDs are very different objects and we discuss this
point in detail in $\S$~\ref{sec:methods}.

The relatively low luminosities of WDs and BDs place their IHZs very
close to the primary, $\sim 0.01$ AU. At these distances tidal effects
are important, and tidal heating may provide enough surface energy to
drive a runaway greenhouse \citep{Barnes12_TV}. Therefore planets
orbiting WDs and BDs must also avoid desiccation via the ``tidal
greenhouse.'' As planetary tidal heating scales with primary mass, WDs can tidally heat
planets more effectively than BDs, all other things being equal.

Beyond identifying planets in danger of desiccation, we also
explore the diversity of terrestrial exoplanets orbiting WDs and BDs.
In and around the IHZ, radiative and tidal heating will produce a mix
of planetary properties, as heating fluxes can be lower than
$F_{crit}$. For example, some planets in the IHZ may only be tidally
heated as much as Io
\citep{Jackson08_hab,Barnes09_THZ,Heller11}. Therefore, if planets are
found around these objects, and within the IHZ, they may be
categorized based on the strengths of tidal and radiative
heating. We find a wide array is possible, including several
types of Venus-like planets, and some exotic types of habitable
planets that do not exist in the Solar System.

\cite{Bolmont11} considered the tidal evolution of the
orbits of planets around BDs, but largely ignored tidal
heating. The orbital evolution can be quite complicated as some
orbits will shrink while others expand, depending on the spin period
of the BD and the orbital semi-major axis. They found that planets may
spend a short amount of time in the IHZ due to tidal evolution.  In
this study, we examine how the cooling of the primary and the tidal
heating of a planet orbiting a WD or BD affect the likelihood of
habitability. We find that the radiation environment prior to arrival in the IHZ is at least as important as the orbital evolution.

This paper is organized as follows. In $\S$~\ref{sec:methods}, we
outline models for primary cooling, the IHZ, tidal heating, hydrogen
loss, and
describe a planetary classification scheme based on tidal and
radiative heating. In
$\S$~\ref{sec:wd}, we examine planets orbiting WDs. In
$\S$~\ref{sec:bd} we turn to BDs. Finally, in
$\S$~\ref{sec:concl}, we draw conclusions.

\section{Methodology\label{sec:methods}}

This study relies heavily on results from
\cite{Barnes12_TV}, \cite{Agol11},
and \cite{Bolmont11}. We first describe the cooling properties of WDs
and BDs and the IHZ. Next we combine them to show how the IHZs evolve
with time. We briefly review tidal phenomena but do not repeat all the
equations here, as they are presented in their entirety in
\citep[][App.~D]{Barnes12_TV}. Finally, we describe a planetary
classification scheme that
categorizes planets based on their insolation and tidal heat flux at
the surface.

\subsection{White Dwarf Cooling}

\begin{figure}
\plotone{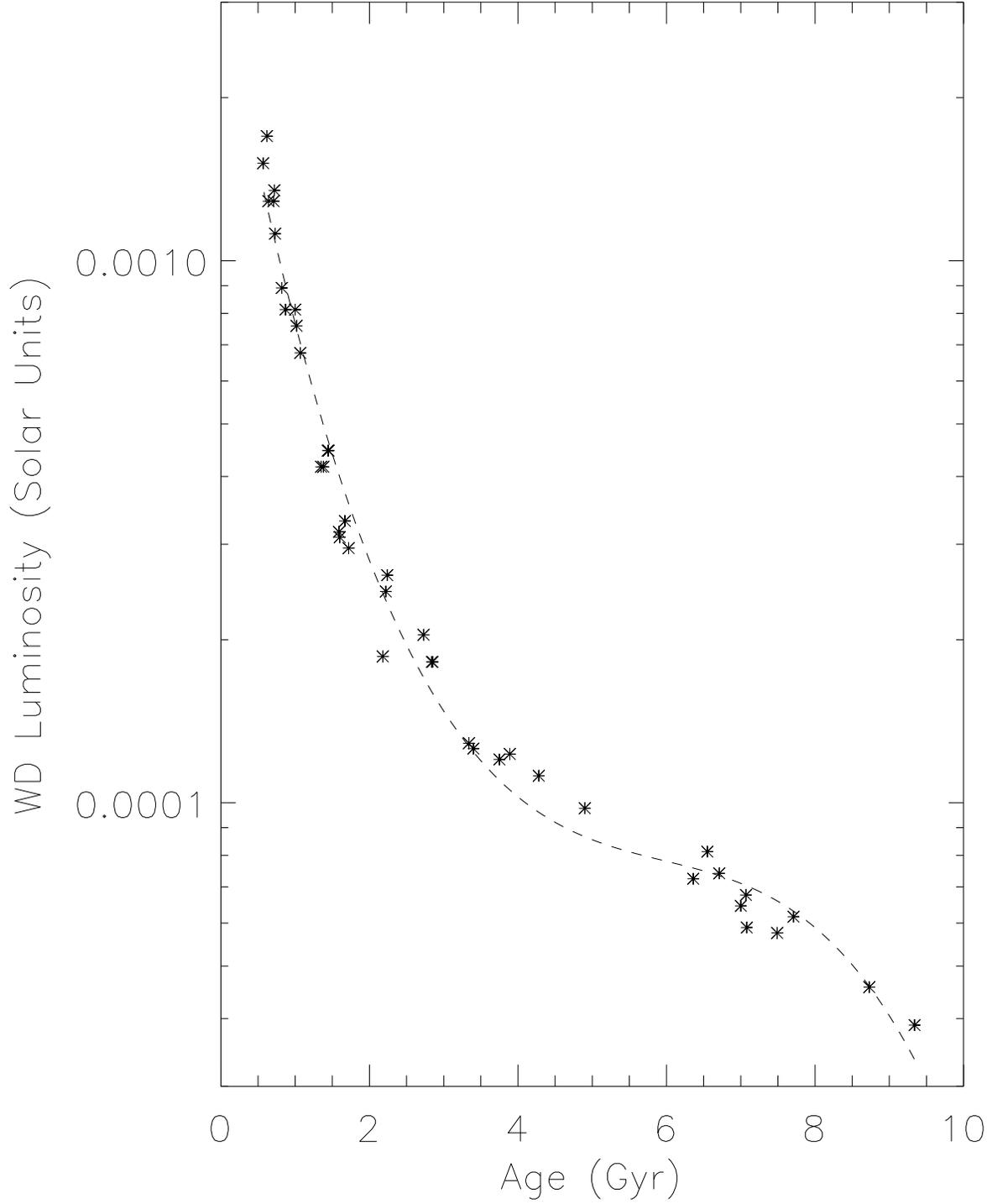}
\caption{\label{fig:wdcool}Luminosity as a function of cooling time
for the WDs in the \cite{Bergeron01} survey with masses
$0.55 \le M_* \le 0.65~\msun$ (stars), and the analytic fit of
Eq.~\eqref{eq:wdlum} (dashed curve).}
\end{figure}

Most WDs have masses near 0.65 $\msun$ with a dispersion of
$0.2~\msun$ \citep{Bergeron01}. \citet{Agol11} focused on $0.6~\msun$
WDs, so we do the same here. \citet{Bergeron01} compute masses, ages
and effective temperature for 152 white dwarfs using theoretical
models and we excised all WDs with masses $<0.55~\msun$ and $>0.65~\msun$
from their Table~2 and plot the remaining 41 objects as stars in
Fig.~\ref{fig:wdcool}. 

Next we fit a third-order polynomial to the data (assuming uniform
uncertainties per point) in order to produce a WD cooling function. If
$L$ is the  base-10 logarithm luminosity in solar units and $t$ the time in Gyr, then we find
\begin{equation}
\label{eq:wdlum}
L = -2.478 - 0.7505t + 0.1199t^2 - 6.686\times 10^{-3}t^3.
\end{equation}
WDs cool rapidly for about
3~Gyr, then level off before falling off again at about 7~Gyr. The
relatively constant luminosity from 3 -- 7~Gyr is probably due to the
crystallization of the degenerate core
\citep{Salpeter61,Segratain94,Hernanz94,Metcalfe04}. Note that the age
given on the abscissa of Fig.~\ref{fig:wdcool} is the WD cooling time only.
It does not include the pre-WD age of the former main-sequence star, so we
neglect the pre-WD circumstances of the system.

\subsection{Brown Dwarf Cooling}

\begin{figure}
\plotone{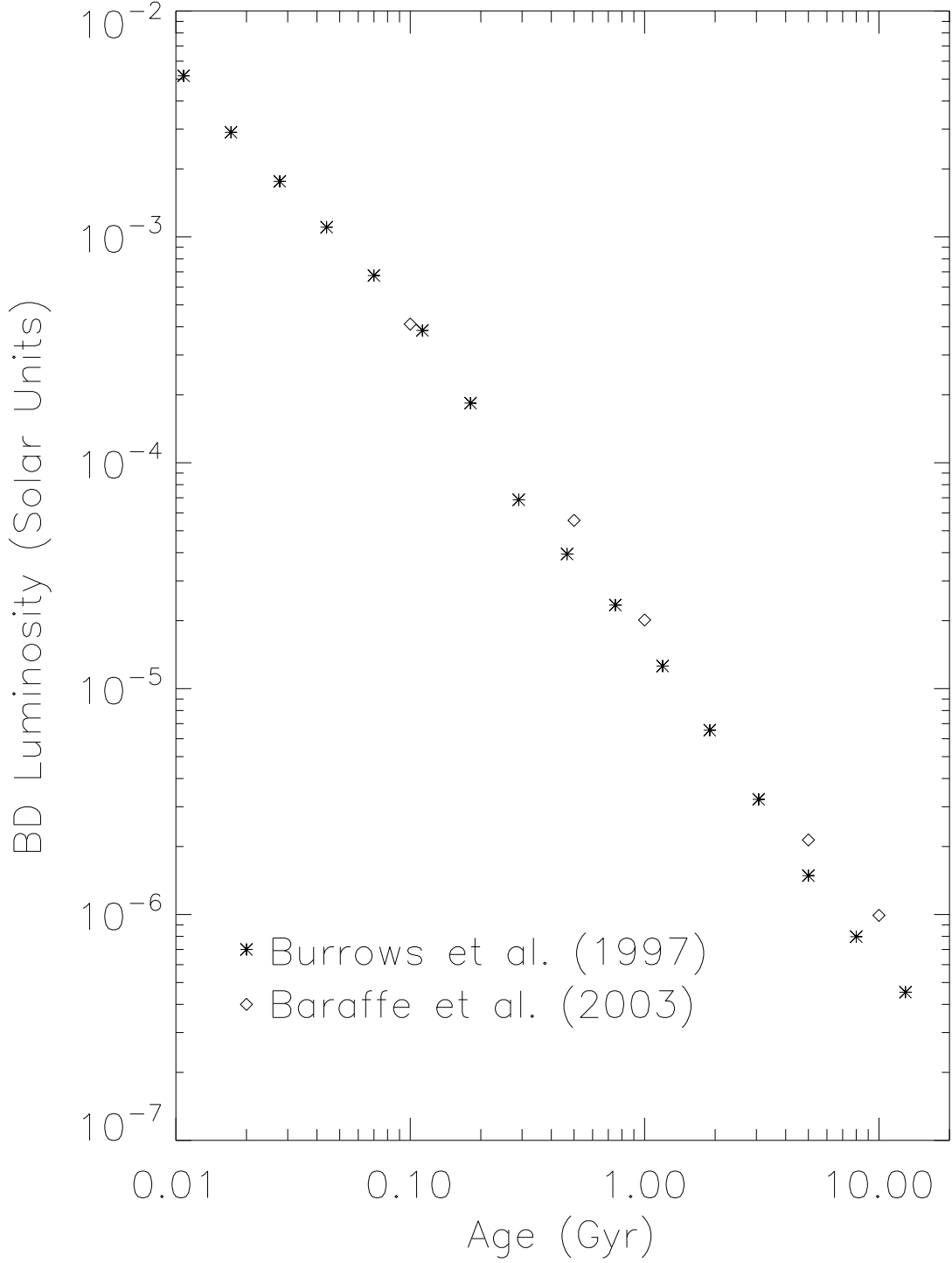}
\caption{\label{fig:bdcool}Luminosity as a function of time for a 42
Jupiter mass
BD according to \cite{Burrows97} (stars) and \cite{Baraffe03} (diamonds).}
\end{figure}

BDs are objects that burn deuterium, but not hydrogen, and are
thus less luminous than MS stars. Traditionally, BDs are assumed to
have masses in the range 13 -- 80 $\mjup$, but the actual limits are
more complicated \citep[see][for a review]{Spiegel11}. Most of their
luminosity results from the release of gravitational energy via
contraction. As relatively few BDs are known and ages are difficult to
constrain empirically, we rely on theoretical models to determine
their luminosity $L_{BD}$ and effective temperature $T_{eff}$. Two
standard models are available \citep{Burrows97,Baraffe03} and their
results are shown in Fig.~\ref{fig:bdcool}. The two models agree with
each other, and we use \cite{Baraffe03} in order to be consistent with
\cite{Bolmont11}.

\subsection{The Insolation Habitable Zone}

We follow exactly the assumptions, models and symbols presented in
\cite{Barnes12_TV}. We label the locus of orbits for which starlight can provide the appropriate level of energy for surface water the IHZ.  We use the \cite{Barnes08_hab} IHZ that merges the work of
\cite{Selsis07} and \cite{WilliamsPollard02} in order to estimate IHZ
boundaries for eccentric orbits and ignore the role of obliquity \citep[\cf][]{Dressing10}

With the previous assumptions, the inner edge of the IHZ is located at
\begin{equation}
\label{eq:lin}
l_{in} = (l_{in\odot} - a_{in}T_* - b_{in}T_*^2)\Big(\frac{L_*}{L_\odot}\Big)^{1/2}(1-e^2)^{-1/4},
\end{equation}
and the outer edge at
\begin{equation}
\label{eq:lout}
l_{out} = (l_{out\odot} - a_{out}T_* - b_{out}T_*^2)\Big(\frac{L_*}{L_\odot}\Big)^{1/2}(1-e^2)^{-1/4}.
\end{equation}
In these equations $l_{in}$ and $l_{out}$ are the inner and outer
edges of the IHZ, respectively, in AU, $l_{in\odot}$ and
$l_{out\odot}$ are the inner and outer edges of the IHZ in the solar
system, respectively, in AU, $a_{in} = 2.7619 \times 10^{-5}$ AU/K,
$b_{in} = 3.8095 \times 10^{-9}$ AU/K$^2$, $a_{out} = 1.3786 \times
10^{-4}$ AU/K, and $b_{out} = 1.4286 \times 10^{-9}$ AU/K$^2$ are
empirically determined constants, and $L_*$ and $L_\odot$ are the
primary's and solar luminosity, respectively.  $T_* = T_{eff} -
5700$\,K, where $T_{eff}$ is the ``effective temperature'' of the
primary
\begin{equation}
\label{eq:Teff}
T_{eff} = \big(\frac{L_*}{4\pi\sigma R_*^2}\big)^{\frac{1}{4}},
\end{equation}
where $R_*$ is the primary's radius.

\begin{figure}
\plotone{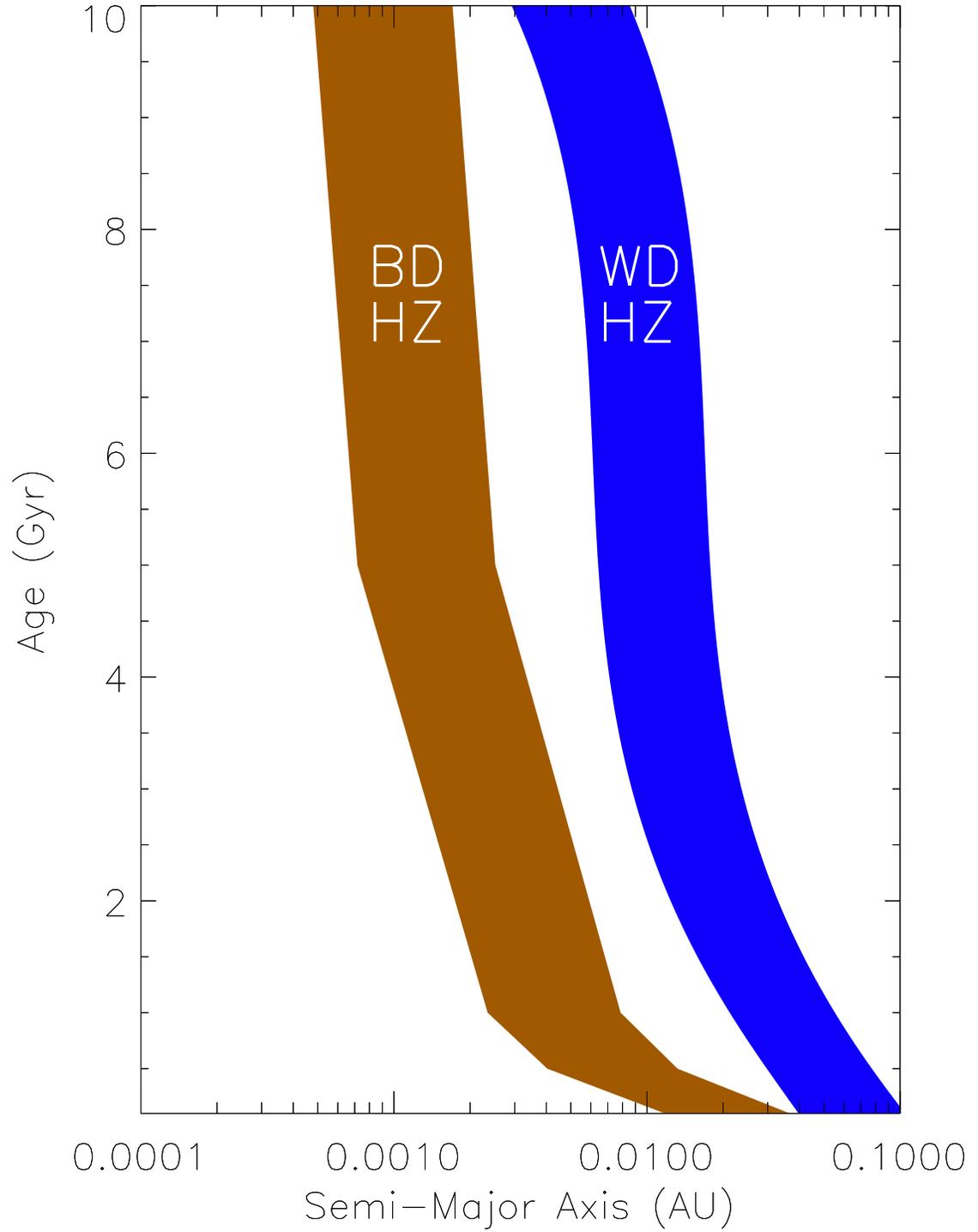}
\caption{\label{fig:hzevol}Evolution of the locations of the IHZ for a $40~\mjup$ BD (brown) and $0.06~\msun$ WD, assuming the \cite{Baraffe03} and \cite{Bergeron01} cooling models, respectively.}
\end{figure}

We can combine the cooling rates with the IHZ models to determine how
the position of the IHZ evolves with time. In
Fig.~\ref{fig:hzevol} we show the location of the IHZs in time for the $0.06~\msun$ WD and $40~\mjup$ BD cases described previously. Initially the BD's IHZ is only a factor of 2 closer in than the WD's, but for ages greater than 1 Gyr, the BD IHZ is about order of magnitude closer to the host.

\subsection{Spectral energy distributions of white dwarfs and brown dwarfs}

\begin{figure*}[t]
  \centering
  \scalebox{0.55}{\includegraphics{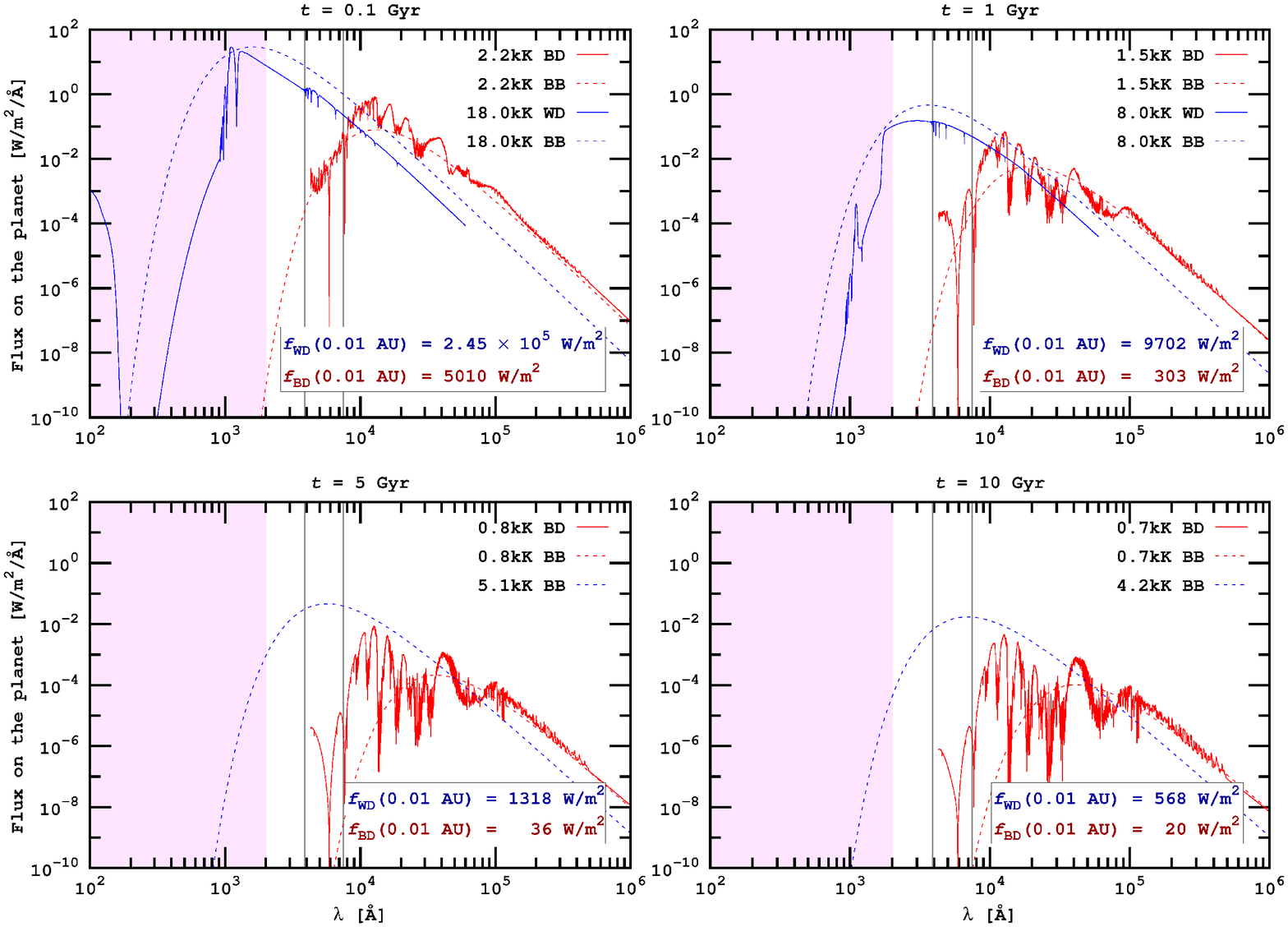}}
  \caption{Evolution of the spectra of a $0.6\,M_\mathrm{Sun}$ WD and
  a $40\,M_\mathrm{J}$ BD. The XUV range, corresponding to those
  wavelengths that may photolyze water vapor, is shaded pink, while vertical lines
denote the visible range used for photosynthesis on Earth. In the
  lower two panels we do not plot the corresponding WD spectra because
  such extremely low temperature models are not available. We also show the Planck curve of a WD-sized and a BD-sized
  black-body (BB) at $0.01$\,AU, respectively. } \label{fig:spectra}
\end{figure*}

Not only will the luminosities of cooling WDs and BDs change, but so
will their spectral energy distribution. The evolution of the XUV
portion of the spectrum (1 -- 2000\AA) is particularly relevant, as photons with this energy
level are the most effective at photolyzing water vapor
\citep{Watson81}, see below. Moreover, the evolution of the spectral energy distribution is important
against the background that almost all land-based life on Earth
depends on photosynthesis, which operates almost exclusively in the
visible range, i.e.  between 400~nm and 700~nm. WD spectra typically
peak in the UV, while BD emission is mainly in the infrared
(IR). Thus, we shall now treat the evolution of WD and BD spectra.

We begin with the WD and consider its spectrum at
$0.1$, $1$, $5$, and $10$\,Gyr. Therefore, we look up the surface
gravity ($\log(g_\mathrm{WD})$), effective temperature
($T_\mathrm{eff,WD}$), and radius ($R_\mathrm{WD}$) of a
$0.609\,M_\mathrm{Sun}$ WD in the $Z~=~0.01$ evolution tables from
\citet{Renedo10}, $Z$ being the progenitor metallicity. We find that
$\log(g_\mathrm{WD})$ increases only slightly from $7.98$\,dex to
$8.05$\,dex, owed to the shrinking from $0.0132$ to
$0.0122\,R_\mathrm{Sun}$. Meanwhile, the effective temperature
decreases from $18,400$\,K to $4,200$\,K. In Fig.~\ref{fig:spectra}
we plot the corresponding WD synthetic spectra
\citep{Finley97,Koester10}
weighted by $(R_\mathrm{WD}/d)^2$, where $d~=~0.01$\,AU is the distance of a
potential planet from the WD. At $5$ and $10$\,Gyr the WD has cooled to effective
temperatures for which we do not have the synthetic spectra available. Thus, we
plot the corresponding black-body profile, from which the WD would deviate
less than in the upper two panels, as the WD does not emit much in the
XUV.

We next consider the BD spectra corresponding to the same
epochs. We choose a $40\,M_\mathrm{Jup}$ BD and look up the effective
temperatures ($T_\mathrm{eff,BD}$) and radii ($R_\mathrm{WD}$) in
\citet{Baraffe03}. We then use cloud-free BD spectral models from
\citet{Burrows06} at $T_\mathrm{eff,BD}~=~2,200$\,K,
$1,500$\,K, $800$\,K, and $700$\,K\footnote{available at
\url{www.astro.princeton.edu/$\sim$burrows}} and weight them with a factor
$(R_\mathrm{BD}/d)^2$, analogous to the procedure for the WD.

Figure~\ref{fig:spectra} finally displays the WD and BD spectral flux
received at a distance of $0.01$\,AU at $0.1$, $1$, $5$, and
$10$\,Gyr. We indicate the XUV range (shaded in pink) to emphasize
those conditions that are most likely to remove surface and
atmospheric water. While the WD spectrum at ages $>~1$\,Gyr becomes
more and more like the Sun's, the BD departs far into the IR. The
bolometric flux of the hosts (WD and BD) at a distance of $0.01$\,AU
is given in the lower right corner of each panel. Recall that the
solar constant, i.e. the Sun's flux at $1$\,AU, is roughly
$1400\,\mathrm{W/m}^2$.
 
\subsection{The Desiccation Timescale}

As mentioned above, XUV photons are able to liberate hydrogen atoms
from terrestrial planets, a process that removes water. If the XUV
flux is high enough and lasts long enough, a planet can become
desiccated and inhospitable for life. Crucially, the water must
be photolyzed first (by UV photons with wavelengths of 120--200 nm),
in order to produce the hydrogen atoms that can escape. While WDs are
initially very hot and have a peak brightness in this regime (see
Fig.~\ref{fig:spectra}), BDs are relatively cool and  UV and XUV
radiation must come from emission due to activity. Observations of
young BDs at these wavelengths are scarce, and hence we cannot
determine the likelihood that planets interior to the HZ will in fact
lose their water.  In this section, we review a simple model for
hydrogen loss, parametrized in terms of XUV luminosity $L_{XUV}$ and
the efficiency of converting that energy into escaping hydrogen atoms.

 First, we must recognize that a wide range of water fractions are
possible for terrestrial exoplanets
\citep[\eg][]{Morbidelli00,Leger04,Raymond04,Bond10}. Planets with
more initial water content are obviously better suited to resist total
desiccation, as removal requires longer time. Similarly,
planets with very low water content, such as the ``dry planets''
proposed by \citep{Abe11} would desiccate more quickly than an
Earth-like planet. Here we assume planets
that begin with Earth's current inventory of water, and that all that
the only source of liberated hydrogen atoms in the upper atmosphere is water.

We use the atmospheric mass loss model described in \cite{Erkaev07},
which improves the standard model by \cite{Watson81}. In the Watson et
al. picture, a layer in the atmosphere exists where absorption of high
energy photons heats the particles to a temperature that permits
escape. Photons in the X-Ray and extreme ultraviolet have the energy
to drive this escape on most self-consistent atmospheres. In essence,
the particles carry away the excess solar energy.

\cite{Watson81} calculated that Venus would lose its water in 280~Myr. The actual
value is probably less than that, as they were unaware that XUV
emission is larger for younger stars, as the spacecraft capable of
such observations, such as {\it ROSAT, EUVE, FUSE, and {\it HST}
\citep{Ribas05},}  had yet to be launched. Furthermore, their model did
not consider the possibility of mass-loss through Lagrange points, or
``Roche lobe overflow.'' \cite{Erkaev07} provided a slight
modification to the \cite{Watson81} model that accounts for this
phenomenon on hot Jupiters. Here we use this model for a
terrestrial exoplanet, which is a fundamentally different
object. However, molecules above the Roche lobe should escape
regardless of their composition, and hence we use the Erkaev model. We
stress that the physics of atmospheric escape, especially for close-in
terrestrial exoplanets, is complicated and poorly understood. In the
\cite{Erkaev07} model, mass is lost at a rate of
\begin{equation}
\label{eq:massloss}
\frac{dM_p}{dt} = - \frac{\pi R_x^2R_p\epsilon F_{XUV}}{GM_pk_{tide}},
\end{equation}
where $R_x$ is the radius of the atmosphere at which the optical depth for
stellar XUV photons is unity, $\epsilon$ is the efficiency of
converting these photons into the kinetic energy of escaping
particles, $F_{XUV}$ is the incident flux of the photons. The
parameter $k_{tide}$ is the correction for Roche lobe overflow:
\begin{equation}
\label{eq:ktide}
k_{tide} = 1 - \frac{3}{2\chi} + \frac{1}{2\chi^3} < 1,
\end{equation}
and 
\begin{equation}
\chi = \Big(\frac{M_P}{3M_*}\Big)^3\frac{a}{R_x}
\end{equation}
is the ratio of the ``Hill radius'' to the radius at the absorbing
layer. From Eq.~(\ref{eq:massloss}) it is trivial to show that 
\begin{equation}
\label{eq:tdes}
t_{des} = \frac{Gm_pM_Hk_{tide}}{\pi R_x^2R_p\epsilon F_{XUV}},
\end{equation}
where $M_H$ is the total mass of all the hydrogen atoms that must be
lost ($\sim 1.4 \times 10^{-5}~\mearth$ in our model, corresponding to
$10^{-4}~\mearth$ of initial water) and assuming $F_{XUV}$ is constant. For the cases we
consider, the additional mass loss via Roche lobe overflow is $\sim 1$\%.

The values of $F_{XUV}$ and $\epsilon$ are therefore the key
parameters as they represent the magnitude and efficiency of the
process. As not all the absorbed energy removes particles, $\epsilon$
must be less than 1. Observations of hot Jupiters suggest values of
$\epsilon$ of ~0.4 \citep{Yelle04,Lammer09,Jackson09}, while on Venus
$\epsilon \sim 0.15$ \citep{Chassefiere96}. However, our model also
requires these photons to dissociate the water molecules, hence we
should expect $\epsilon$ in a water-rich atmosphere to be much less than on a
hot Jupiter with a predominantly hydrogen atmosphere. For reference,
using the assumptions of \cite{Watson81}, we find $\epsilon = 1.7
\times 10^{-4}$ implies $t_{des} = 280$~Myr for Venus. Note that a
water-rich planet that loses its hydrogen may develop an oxygen-rich
atmosphere and may be misidentified as potentially
habitable. Furthermore, the destruction of O$_3$ by UV radiation would
also be slow, and hence could lead to a ``false positive'' for life
\citep{DomagalGoldman12}. This possibility requires detailed
photochemical modeling for confirmation, and was beyond the scope of this study.

\begin{figure}
\includegraphics[width=5in]{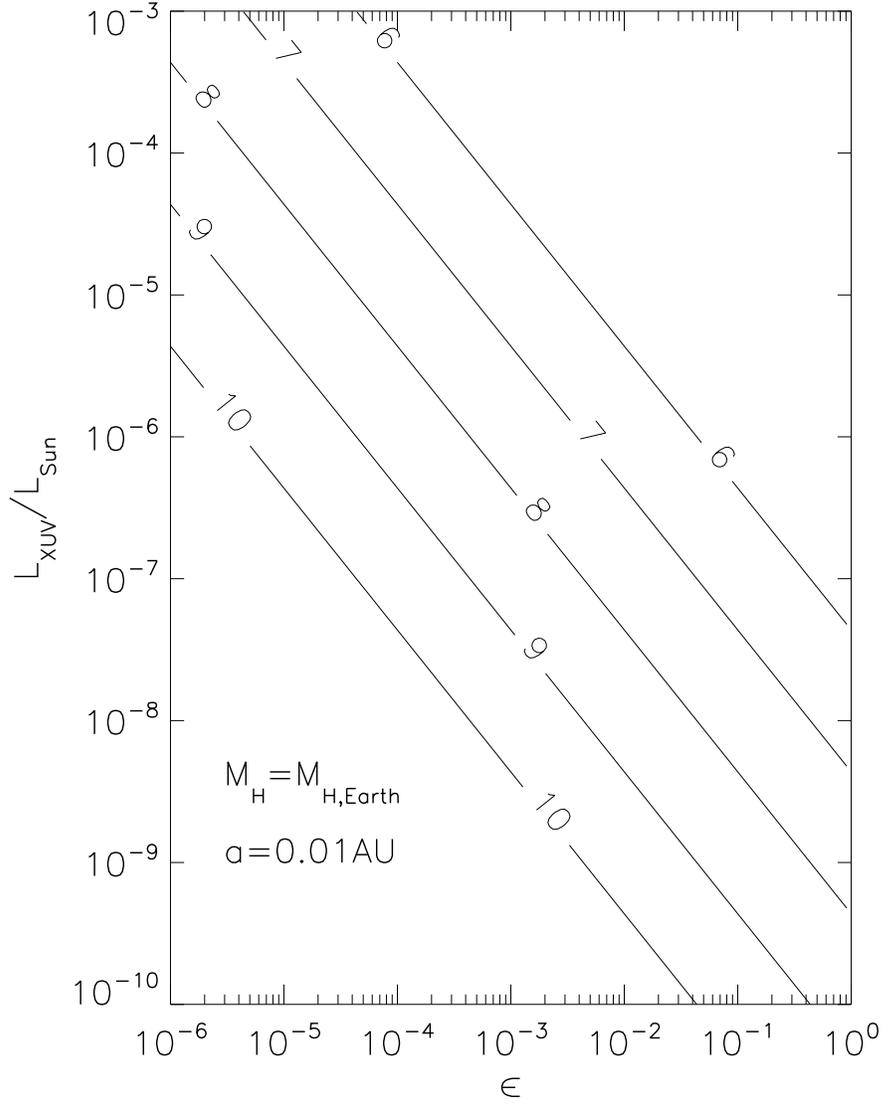}
\caption{\label{fig:tdes} Desiccation timescale for an Earth-like
planet orbiting a BD or WD at 0.01~AU. Contour lines represent the
logarithm of the time for the Earth's inventory of hydrogen to be
lost, \ie the planet's mass of hydrogen in water $M_H$ is equal to the
Earth's $M_{\textrm{H,EARTH}}$.}
\end{figure}

In Fig.~\ref{fig:tdes} we show the desiccation timescale for an
Earth-like planet orbiting a WD or BD at 0.01~AU. This distance is
just interior to the IHZ at 100~Myr, and is therefore a critical
location for a planet to experience sustained habitability in the
future. For most young WDs, the peak brightness lies in the XUV, and
the total luminosity is $>10^{-3}~\lsun$. Thus, their planets are in
an environment characterized by the top portion of the graph. In order
to avoid desiccation, those planets must have $\epsilon <
10^{-6}$. This is a very small value, and we expect that planets
initially interior to the IHZ will be desiccated by the time it
arrives.

The situation for BD planets is difficult to assess at present.
The two steps of photolysis and escape require knowledge of the
evolution of UV and XUV emission, which is largely unknown at
present. The cool temperatures of BDs suggest that photolysis may be
the limiting step in the desiccation process, as the BD photospheres
may be too cool to emit enough UV radiation to drive
desiccation. Assuming that water molecules can be destroyed, it is
unclear if the XUV emission is large enough to drive mass loss,
too. We can appeal to low-mass M dwarfs for guidance and extrapolate
the results of \cite{Pizzolato03} to the BD realm. They estimated the
``saturation level'' of XUV radiation, which is the largest amount of
XUV luminosity as a fraction of total luminosity, observed in
stars. They find that for stars with masses in the range $0.22 -
0.6~\msun$ that the XUV emission never represents more than $10^{-3}$
of the total. At 100 Myr, BDs are expected to have luminosities near
$10^{-4}~\lsun$, and hence their XUV flux should be less than
$10^{-7}~\lsun$. In order for a planet initially at 0.01~AU from a BD
to avoid desiccation, $\epsilon < 5 \times 10^{-3}$. We caution that
BD interiors are expected to be fully convective, while stars in the
above mass range are not, and hence the stars in \cite{Pizzolato03} may
not be analogous to BDs. A survey of XUV emission from a large sample
of BDs is sorely needed to address this aspect of planetary
habitability, but will be very challenging, even from space. On the
other hand, spectral characterization of atmospheres in the IHZs of
older BDs could provide the tightest constraints on the early
high-energy emission of BDs.
 
\subsection{Tidal Effects}

We use the two tidal models described in \cite{Heller11}, and the
numerical methods described in \citep{Barnes12_TV}, see also \citep{FerrazMello08,Greenberg09,Leconte10}. These
models are qualitatively different, but widely used in astrophysics and
planetary science. They both assume that the tidal bulge
raised on a body lags behind the apparent position of a perturber
and that the shape of the deformed body can be adequately modeled as a
superposition of surface waves. One, which we call the
``constant time lag'' (CTL) model, assumes that the time between the
passage of the perturber and the passage of the tidal bulge is
constant. The tidal effects are encapsulated in the product of the
``tidal time lag'' $\tau$ and the second-order tidal Love number $k_2$.
The other model, which we call the ``constant phase lag''
(CPL) model, assumes that the angle between the lines connecting the centers
of the two bodies and the center of the deformed body to the bulge is
constant. The tidal effects in this model lie in the quotient of the
``tidal quality factor'' $Q$ and $k_2$. The tidal time lag and quality factor
essentially measure how effectively energy is dissipated inside a body.

As in \cite{Barnes12_TV}, we assume the planets are
Earth-like with $\tau = 640$~s or $Q = 10$ \citep{Lambeck77,Yoder95}.
The tidal heating scales linearly with these parameters and as semi-major
axis $a$ to the -7.5, eccentricity $e$ squared, obliquity $\psi_p$
squared and spin frequency $\omega_p$ squared, \citep[see][]{Barnes12_TV}. The heat flux $F$ through the
surface is critical for surface life, and is also a measure of how
geophysical processes transport energy. Martian tectonics is
believed to have shut down at a heat flux of
0.04~\wpmsq~\citep{Williams97}, The Earth's heat flux, due to
non-tidal processes, is 0.08~\wpmsq~\citep{Pollack93}, Io's heat flux
due to tidal heating, is 2~\wpmsq~\citep{Strom79,Veeder94}, and the
limit to trigger a runaway greenhouse is $F_{crit} =
300$~\wpmsq~\citep{Pierrehumbert10}. We will use these
limits to
explore the habitability of planets orbiting WDs and BDs (see
below). We shall thereby keep in mind that these values are not
natural or material constants but depend strongly on a planet's
composition, mass, and age.

The tidal Q of WDs is poorly constrained. Recently, \cite{Piro11} used
the luminosities of an eclipsing He-C/O WD binary to place limits on
the Q values of the two companions. He found the former has $Q < 7
\times 10^{10}$ and the latter has $Q < 2 \times 10^7$. Unfortunately
these are relatively rare classes of WDs, as 80\% are classified as H
WDs. These values are probably larger than for solar-type
stars \citep{Jackson09}, and predict little dissipation in the WD from
the terrestrial planet. Furthermore, these values preclude the
possibility that the planet's orbit could tidally decay at a rate that
allows it to remain in the IHZ as the WD cools. At 100 Myr, the IHZ is
at $\sim 0.05$~AU, and the tide on the WD is negligible. This assumes
the WD rotation period is longer than the orbital period, which may
not be the case. The rotation rates of WDs can span a wide range, from
seconds to years, but should initially rotate with periods longer than
one day \citep{Kawaler04,Boshkayev12}.

Tidal processes in BDs are also relatively unexplored. \cite{Heller10}
used the CPL and CTL models to explore the range of tidal heating
available in the eclipsing BD binary pair 2MASS J05352184-0546085
\citep{Stassun06}. This study indicates that tidal heating can be
significant in BDs. Tidal evolution of BDs orbiting main sequence
stars has also been calculated for a few cases,
\cite[\eg][]{Fleming10,Lee11}, but tidal dissipation in BDs remains
poorly constrained. Here we assume $Q = 10^6$, which is a reasonable
estimate based on those prior studies. We use the radius values as a
function of time given in \cite{Baraffe03}.

\subsection{Classifying Planets by Radiative and Tidal Heat}

Different levels of insolation and tidal heat can be used to devise a
classification scheme for terrestrial exoplanets orbiting in and
around the IHZ. Here we propose categories that are relevant for
the discussions in $\S\S$~\ref{sec:wd}--\ref{sec:bd}. The categories
are defined by the origin and total amount of upward energy flux at
the planetary surface. That due to orbit-averaged insolation,  and
assuming efficient energy redistribution, is
\begin{equation}\label{eq:insolation}
F_{insol} = \frac{L_p}{16\pi a^2\sqrt{1-e^2}}(1-A),
\end{equation}
\noindent
where $L_p$ is the primary's luminosity, $a$ is semi-major axis, $e$
is eccentricity, and $A$ is the planetary albedo. In addition to
this heating, there is also an energy flux resulting from the cooling
of the planetary interior. Three sources are known to be possible:
radioactive decay, latent heat of formation, and tides. Here we will
focus on the tidal heat. The amount of radioactive and latent heat is
very difficult to calculate from observations, whereas the tidal heat
can at least be scaled by a single parameter. When the tidal heat
drops below a level that is geophysically interesting, then we will
assume that the planet's interior behaves like the Earth, which has a
heat flux of 0.08~\wpmsq, of which a negligible fraction is due to tides.  We are
therefore assuming that in the absence of tidal heating, the planetary
interior behaves like the Earth's.  The flux due to tides is
$F_{tide}$ \citep[see][]{Barnes12_TV}. The sum of $F_{insol}$ and
$F_{tide}$ is $F_{tot}$.

Three Venus-like planets are possible. An ``Insolation Venus'' is a
planet that receives enough stellar radiation to trigger a moist or
runaway greenhouse.  A ``Tidal Venus'' is a planet whose tidal heat
flux exceeds the runaway greenhouse flux, $F_{tide} \ge
F_{crit}$. Finally, ``Tidal-Insolation Venuses'' are planets with
$F_{tide} < F_{crit}$ and $F_{insol} < F_{crit}$, but $F_{tide} +
F_{insol} \ge F_{crit}$. 

``Super-Ios'' are worlds that experience tidal heating as large or
larger than Io's ($F_{tide} \gsim 2$~\wpmsq), but less than $F_{crit}$
\citep{Jackson08_hab}, see also
\cite{Barnes09_40307,Barnes09_THZ}. We assume
that the dissipation in the interior determines the onset of Io-like
volcanism and that tidal dissipation in the interior is 10
times less effective than in an ocean. Therefore, we determine the
Super-Io boundary using $\tau = 64$~s or $Q = 100$.

In the IHZ are two types of habitable planets. Planets with
2~\wpmsq~$\ge~F_{tide}~\ge~$ 0.04~\wpmsq~are ``Tidal Earths.'' Those
planets with $F_{tide} < 0.04$~\wpmsq~are ``Earth Twins'' as they are
in the IHZ and receive negligible tidal heating.

Planets exterior to the IHZ and with $2 < F_{tide} < 0.04$~\wpmsq~are
labeled ``Super-Europas,'' while those with $F_{tide} <
0.04$~\wpmsq~are ``Snowballs.'' While both may possess layers of
sub-surface water, these environments are not detectable across
interstellar distances.

%
%
%
%

\section{Planets Orbiting White Dwarfs \label{sec:wd}}

\subsection{$0.6~\msun$ White Dwarfs} 

\begin{figure}
\plotone{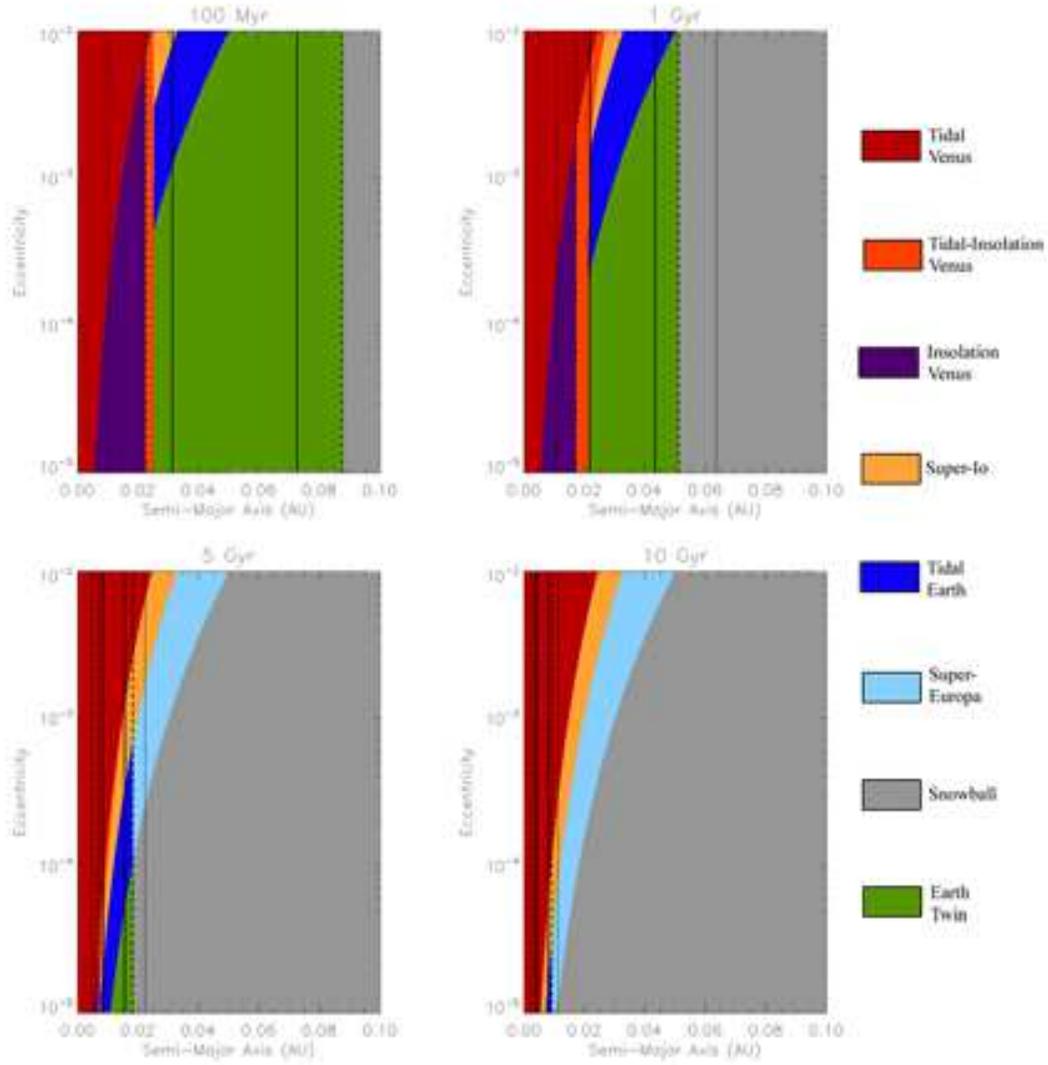}
\caption{\label{fig:wd}Planetary classifications for a tidally-locked $1~\mearth$
planet with no obliquity in orbit about a $0.6~\msun$ WD. As WDs cool significantly with time, we show the
phases and IHZ as a function of time, from 100 Myr (top left) to 10
Gyr (bottom right).}
\end{figure}

The luminosity of a $0.6~\msun$ WD is $\sim 10^{-3}~\lsun$
\citep{Bergeron01}, and hence the IHZ is located at $\sim 0.01$ AU
\citep{Agol11}. In Fig.~\ref{fig:wd} we show the planetary classifications
of $1~\mearth$
planets in orbit around a $0.6~\msun$ WD at 4 different times, assuming
the cooling model derived in $\S$~\ref{sec:methods}. Here we assume that
the planets rotate with the equilibrium period and with no obliquity. Note
the low values of eccentricity, a testament to the power of tides in these
systems.

At 100 Myr, the WD is very hot, $\sim 10^4$ K. This effective
temperature is outside the range considered by \cite{Selsis07}, and
hence in order to obtain a close match between the 50\% cloud cover
boundary of \cite{Selsis07} and the runaway greenhouse limit of
\cite{Pierrehumbert10} we set the planetary albedo to
0.87. (For the later times we used $A = 0.5$.)
Even at that large value, the inner boundary of the IHZ does not
align with Pierrehumbert's prediction. Earth Twins require $e \lsim
0.01$.

From 0.1 -- 1 Gyr, the WD cools rapidly, see Fig.~\ref{fig:wdcool},
and afterward the IHZ has moved in by 50\%. Tidal effects desiccate
more of the IHZ at ``large'' eccentricity, and much is in the Tidal
Earth regime. However at 1 Gyr, most of the IHZ below $e = 0.01$ still
appears habitable.

By 5 Gyr, however, (lower left panel), even at very low
eccentricities, most planets in the IHZ will experience strong tidal
effects. At $e = 10^{-5}$, the inner edge is experiencing at least
Tidal Earth conditions and Earth Twins must be at the outer edge and
with $e < 10^{-4}$. \cite{Agol11} identified 4 -- 7 Gyr as a ``sweet
spot'' for planets around WDs, as the cooling levels off, see
$\S$~\ref{sec:methods}, but in order to be habitable, candidate
planets require very low eccentricities in order to avoid a tidal
greenhouse. Note that at $a = 0.01$~AU and $e = 10^{-4}$ the
difference between closest and farthest approach from the host is $2ae
= 300$~km, or about 5\% the radius of the Earth.

At 10 Gyr (lower right panel), the situation has continued to
deteriorate for habitable planets. Not only has the width of the IHZ
shrunk considerably, but Earth Twins are only possible if $e <
10^{-5}$!  Habitability at this stage probably requires a system
consisting of one WD and one planet, as additional planet mass
companions can raise $e$ above this threshold, see below and
\cite{Barnes10_corot7}. As with CoRoT-7~b, the
galactic tide or passing stars cannot pump $e$ to this ``large''
value~\citep{Barnes10_corot7}.

Tides on planets orbiting WDs are a strong constraint on habitability,
but are less important than the WD's cooling. Consider the left panels
of Fig.~\ref{fig:wd}. In the bottom panel, the IHZ lines up with
Insolation Venus (purple) and Tidal Venus (red) regions of the upper
panel. Therefore, planets located in the 5 Gyr IHZ were interior to
the 0.1 Gyr IHZ. Furthermore, young WDs are very hot and emit
substantial energy in the XUV wavelengths that photolyze H$_2$O. As
0.1 Gyr $\sim t_{des}$, we expect planets in the 5 Gyr IHZ to have
been desiccated well before the IHZ reaches them. In other words,
habitability requires special circumstances that override the standard
picture of habitability. Unfortunately tidal decay of the (circular)
orbit is unlikely, as the tidal Q ($\tau$) is probably very large
(small) for WDs
\citep{Agol11,ProdanMurray12}. Therefore, planets orbiting hydrogen WDs will not
spiral in, and cannot keep pace with the shrinking IHZ.

\subsection{KOI 55}

Two planet candidates have recently been proposed that orbit an
extreme horizontal branch star, an astrophysical object that will shortly transition to a WD. These candidates were found with the {\it
Kepler} spacecraft and are designated KOI 55.01 and KOI 55.02
\citep{Charpinet11}. The stellar mass is $0.5~\msun$, and the two planets' radii are 0.76 and $0.87~\rearth$. If Earth-like in composition and structure, these planets have masses of 0.41 and $0.63~\mearth$, respectively \citep{Sotin07}. If these candidates are real, then they will
orbit a WD and hence could become habitable. Moreover, their detection
implies that these planets may be typical of those that orbit WDs. In
this subsection we consider the future habitability of these two
objects.

The two candidates have semi-major axes of 0.006 and 0.0076 AU. The
outer planet, KOI 55.02, therefore falls in the IHZ during the
crystallization epoch from 4--7 Gyr. The primary currently has a
temperature of 27,700~K and therefore emits strongly in the XUV and
X-Ray spectral regimes (see Fig. 3, upper left panel). These planets
are therefore roasting, and likely losing their water (if they had
any). Hence, both objects are in danger of sterilization due to
irradiation now, and are unlikely to become habitable
later. Furthermore, this extremely high temperature suggests our
nominal value for $t_{des}$ may be too high.

There is also a danger that perturbations between the two planets
(which may be in a 3:2 resonance) may maintain a large enough
eccentricity ($\gsim 10^{-4}$) to sustain a tidal greenhouse. To test for this possibility we ran two
numerical simulations of the gravitational perturbations between the
planets using HNBody\footnote{Publicly available at
http://janus.astro.umd.edu/HNBody}. The first case used the nominal
values from \cite{Charpinet11}, and the second forced the two planets
into an exact 3:2 mean motion resonance by pushing the outer planet's
period to 8.64 hours. For the former, the eccentricity of KOI 55.02
oscillated between $5 \times~10^{-6}$ and $5 \times~10^{-5}$. This
range places the planet in the Tidal Earth or Earth Twin regime. For
the latter case, the resonance perturbs the eccentricity more
strongly, and it varies between $10^{-4}$ and
$10^{-3}$. These values are in the Tidal Venus range and the planet
would not be habitable. However, given
the strong tidal forces in this region, these ranges
are probably maximum values. Nonetheless the future habitability of
these two planet candidates seems unlikely.

%
%
%
%

\section{Planets Orbiting Brown Dwarfs \label{sec:bd}}

As in \cite{Bolmont11}, we consider a $0.04~\msun$ BD and employ the
theoretical cooling and contraction models of \cite{Baraffe03}, to
calculate IHZ boundaries. Here we extend the results of
\cite{Selsis07} to the luminosity and temperatures of BDs, which
give similar limits as in \cite{Bolmont11}. In this case, we use the CPL
model with a continuum of rotation states
\citep[see][App.~D.3]{Barnes12_TV} to determine $F_{tide}$, and set
the BD's obliquity to 0 and its rotation period to 1 day (note that
these values do not affect the tidal heating in the planet).

\begin{figure}
\plotone{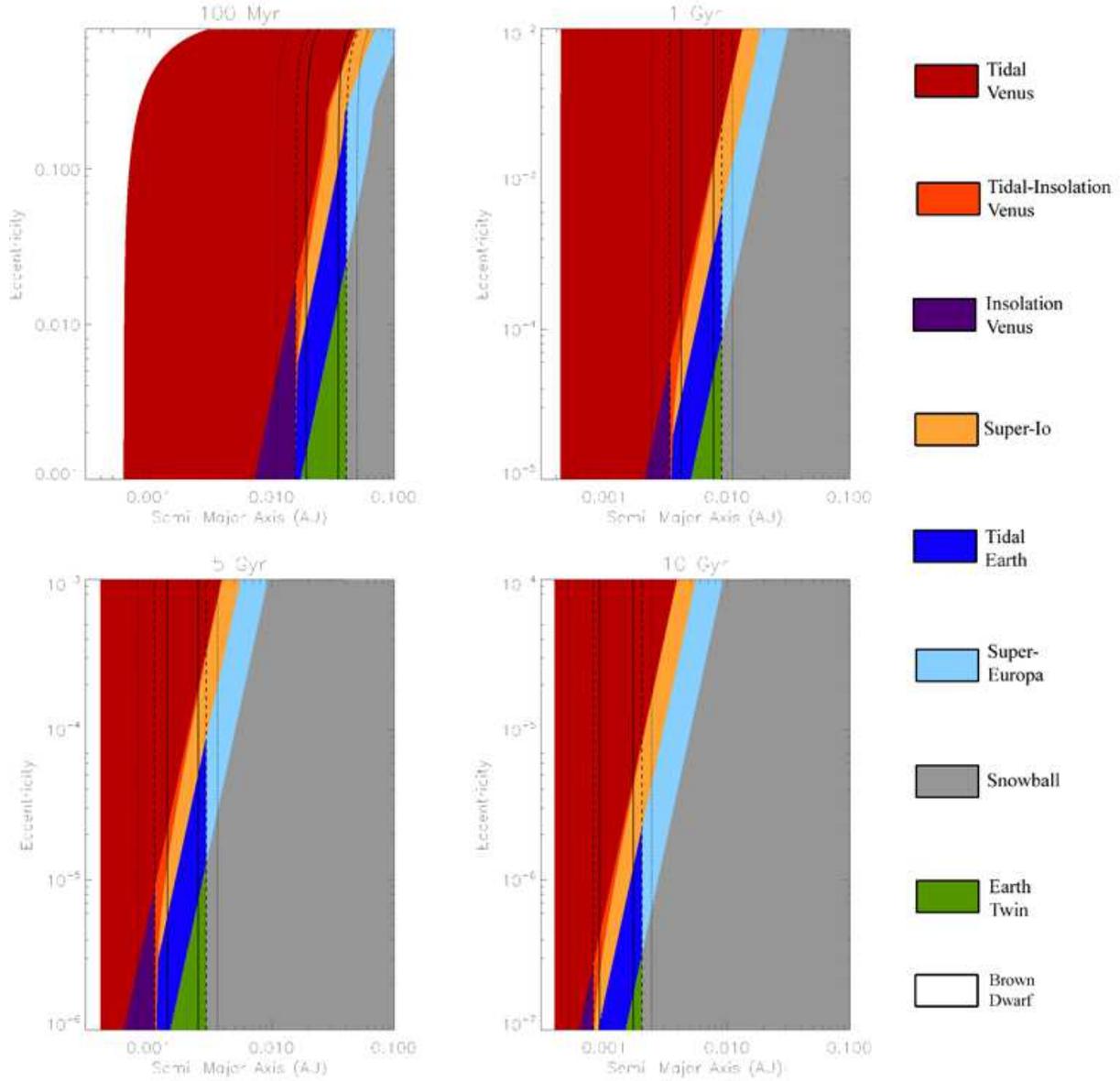}
\caption{\label{fig:bd}Planetary classifications for a tidally-locked $1~\mearth$
planet with no obliquity, in orbit about a $0.04~\msun$ brown
dwarf at 100 Myr (top left), 1~Gyr (top right), 5~Gyr (bottom left), and 10~Gyr (bottom right). White regions correspond to orbits that intersect the BD.}
\end{figure}

In Fig.~\ref{fig:bd} we show the boundaries for planetary classes for
a $1~\mearth$ planet after 100~Myr, 1~Gyr, 5~Gyr and 10~Gyr. Note
that the scales of the axes change for each panel. At 100~Myr, \ie
$t_{des}$, planets in the IHZ with $e > 0.01$ may be in a
Tidal Venus state. Also note that our model predicts that all planets interior to 0.016~AU
will have lost their water. At 1~Gyr planets must have $e < 5 \times
10^{-5}$ to avoid a tidal greenhouse, but at this point, the IHZ has
moved into a region in which all water should already have been
lost. To avoid a Tidal Venus state at 5 and 10~Gyr, eccentricities
must be less than $10^{-6}$ and $10^{-7}$, respectively. Therefore,
even if a planet could maintain its water inventory until the IHZ
reached its orbit, a planet must have an extremely circular orbit to
avoid catastrophic tidal heating. Also note that the Roche lobe,
the distance where a planet is in danger of being torn apart by tides,
lies at $\sim 3 \times 10^{-3}$~AU
\citep{AndreeshchevScalo04,Bolmont11}, further limiting habitability.

\cite{Bolmont11} point out that tides will tend to push planets out, in an analogous
manner as the Moon recedes from the Earth. Therefore, as the planets
move out and the IHZ moves in, the residence time in the IHZ can be
quite short. They considered the example of a planet that begins at
$9\times 10^{-3}$~AU and find that planets between 0.1 and
10~$\mearth$ will require at least 50~Myr to reach the IHZ. As this time is
of order $t_{des}$, the key limit to habitability may be that those
planets will have already lost their water, rather than the relatively
short time the planets spend in the IHZ. However, not enough is known
about activity on young BDs to accurately determine the threat of
photolyzation and hydrogen escape.

%
%
%
%

\section{Conclusions \label{sec:concl}}

Planets orbiting WDs and BDs suffer a number of critical habitability
issues, such as strong tidal heating, water loss due to bombardment by
high energy photons, a spectral energy distribution very
different from the Sun's, and a cooling primary. Here we have
outlined how dire the situation is for these types of planets, using
simple but standard models of the IHZ, tidal heating, and cooling
rates. We conclude that WDs are even less likely to support
habitable planets than BDs due to the former's stronger XUV emission.

As the luminosities of WDs and BDs change dramatically with time, see
Figs.~\ref{fig:wdcool}--\ref{fig:hzevol}, the IHZ moves
inward. Planets interior to the IHZ experience a desiccating
greenhouse phase for times longer than $t_{des}$ due to insolation and
hence lose their water before the IHZ reaches their orbit, and are
hence uninhabitable. This possibility is more likely for WDs than BDs
as the former's peak radiation lies in the near UV, while constraints
on the latter's emission are difficult to obtain. New data on the
duration and strength of BD activity are sorely needed. However, even
if the spectral energy distribution of the host does not remove the
hydrogen, the orbital eccentricities must remain very low in order to
avoid a tidal greenhouse.

Once a planet is in the IHZ, there are four requirements for
habitability: 1) The eccentricity must be low enough to avoid a tidal
greenhouse, 2) a reservoir of water must survive the desiccating
greenhouse phase, perhaps trapped in the mantle, 3) a mechanism must
exist to draw down CO$_2$ and quench the moist greenhouse, and 4) a
(possibly different) mechanism must release the water
reservoir to the surface. Currently, no phenomena are known that can
satisfy all four of these constraints.

On the other hand, the possibility exists that the planet could arrive in the IHZ from a further distance
\citep[see \eg][]{Agol11_ESS}. For example, if a more distant and massive
companion is present, gravitational perturbations could excite the
planet's eccentricity such that the pericenter distance is close
enough for tides to circularize the orbit. Then the planet could
arrive in the IHZ in a circular orbit after tidal circularization
\citep{Nagasawa08}. Alternatively, a late-stage planet-planet
scattering event could move a terrestrial planet from a more distant
orbit to the IHZ. Recent observations suggest that planetary systems
with 2:1 mean motion resonances are younger than average, suggesting
they break apart as they age \citep{KoriskiZucker11}. In both these
scenarios, the planet still has to contend with large eccentricities,
and hence tidal heating could still sterilize the
surface. Additionally, the planet would have to break out of its
frozen state, and it is unclear if such ``cold starts'' are possible
\citep{Kasting93}. Unfortunately these sorts of histories are nearly
impossible to recognize from a planet on a circular orbit, and hence
we should not expect to be able to determine if a planet arrived in
the IHZ without experiencing desiccating conditions. Nonetheless, future work should determine if planets can still be habitable after an extended time in the runaway greenhouse, and the likelihood that planets orbiting WDs and BDs can migrate long after their formation.

Planets are complex objects and we hesitate to rule out habitability,
but planets orbiting BDs and WDs face a difficult path to habitability.
On the other hand, as pointed out by \cite{Agol11},
the signals for some of these objects could be very easily
detected. If a planet is found in the IHZ of a cooling primary,
follow-up observations should still be carried out. Regardless of
their habitability, terrestrial planets found around these objects
will provide fundamentally new insight into planet formation, and
hence into the extent of life in the universe.

\medskip
We thank Victoria Meadows, Kristina Mullins, Jim Kasting, Colin
Goldblatt, Eric Agol, Detlev Koester, David Spiegel and two anonymous
referees for extremely valuable commentary on this concept and
manuscript. RB acknowledges support from NSF grant AST-1108882, and
from the NASA Astrobiology Institute's Virtual Planetary Laboratory
lead team. RH is supported by the Deutsche Forschungsgemeinschaft
(reference number schw536/33-1).

\bibliography{Cooling_AsBio}

\begin{thebibliography}{88}
\expandafter\ifx\csname natexlab\endcsname\relax\def\natexlab#1{#1}\fi

\bibitem[{{Abe}(1993)}]{Abe93}
{Abe}, Y. 1993, Lithos, 30, 223

\bibitem[{{Abe} {et~al.}(2011){Abe}, {Abe-Ouchi}, {Sleep}, \& {Zahnle}}]{Abe11}
{Abe}, Y., {Abe-Ouchi}, A., {Sleep}, N.~H., \& {Zahnle}, K.~J. 2011,
  Astrobiology, 11, 443

\bibitem[{{Agol}(2011{\natexlab{a}})}]{Agol11_ESS}
{Agol}, E. 2011{\natexlab{a}}, in AAS/Division for Extreme Solar Systems
  Abstracts, Vol.~2, AAS/Division for Extreme Solar Systems Abstracts, 1506

\bibitem[{{Agol}(2011{\natexlab{b}})}]{Agol11}
{Agol}, E. 2011{\natexlab{b}}, Astrophys.~J., 731, L31+

\bibitem[{{Andreeshchev} \& {Scalo}(2004)}]{AndreeshchevScalo04}
{Andreeshchev}, A., \& {Scalo}, J. 2004, in IAU Symposium, Vol. 213,
  Bioastronomy 2002: Life Among the Stars, ed. {R.~Norris \& F.~Stootman},
  115--+

\bibitem[{{Apai} {et~al.}(2005){Apai}, {Pascucci}, {Bouwman}, {Natta},
  {Henning}, \& {Dullemond}}]{Apai05}
{Apai}, D., {Pascucci}, I., {Bouwman}, J., {Natta}, A., {Henning}, T., \&
  {Dullemond}, C.~P. 2005, Science, 310, 834

\bibitem[{{Baraffe} {et~al.}(2003){Baraffe}, {Chabrier}, {Barman}, {Allard}, \&
  {Hauschildt}}]{Baraffe03}
{Baraffe}, I., {Chabrier}, G., {Barman}, T.~S., {Allard}, F., \& {Hauschildt},
  P.~H. 2003, Astro.~\& Astrophys., 402, 701

\bibitem[{{Barnes} {et~al.}(2009{\natexlab{a}}){Barnes}, {Jackson},
  {Greenberg}, \& {Raymond}}]{Barnes09_THZ}
{Barnes}, R., {Jackson}, B., {Greenberg}, R., \& {Raymond}, S.~N.
  2009{\natexlab{a}}, Astrophys.~J., 700, L30

\bibitem[{{Barnes} {et~al.}(2009{\natexlab{b}}){Barnes}, {Jackson}, {Raymond},
  {West}, \& {Greenberg}}]{Barnes09_40307}
{Barnes}, R., {Jackson}, B., {Raymond}, S.~N., {West}, A.~A., \& {Greenberg},
  R. 2009{\natexlab{b}}, Astrophys.~J., 695, 1006

\bibitem[{{Barnes} {et~al.}(2012){Barnes}, {Mullins}, {Goldblatt}, {Meadows},
  {Kasting}, \& {Heller}}]{Barnes12_TV}
{Barnes}, R., {Mullins}, K., {Goldblatt}, C., {Meadows}, V.~S., {Kasting},
  J.~F., \& {Heller}, R. 2012, AsBio, submitted

\bibitem[{{Barnes} {et~al.}(2010){Barnes}, {Raymond}, {Greenberg}, {Jackson},
  \& {Kaib}}]{Barnes10_corot7}
{Barnes}, R., {Raymond}, S.~N., {Greenberg}, R., {Jackson}, B., \& {Kaib},
  N.~A. 2010, Astrophys.~J., 709, L95

\bibitem[{{Barnes} {et~al.}(2008){Barnes}, {Raymond}, {Jackson}, \&
  {Greenberg}}]{Barnes08_hab}
{Barnes}, R., {Raymond}, S.~N., {Jackson}, B., \& {Greenberg}, R. 2008,
  Astrobiology, 8, 557

\bibitem[{{Bergeron} {et~al.}(2001){Bergeron}, {Leggett}, \&
  {Ruiz}}]{Bergeron01}
{Bergeron}, P., {Leggett}, S.~K., \& {Ruiz}, M.~T. 2001, Astrophys.~J., 133,
  413

\bibitem[{{Beuermann} {et~al.}(2011){Beuermann}, {Buhlmann}, {Diese},
  {Dreizler}, {Hessman}, {Husser}, {Miller}, {Nickol}, {Pons}, {Ruhr},
  {Schm{\"u}lling}, {Schwope}, {Sorge}, {Ulrichs}, {Winget}, \&
  {Winget}}]{Beuermann11}
{Beuermann}, K. {et~al.} 2011, \aap, 526, A53

\bibitem[{{Beuermann} {et~al.}(2012){Beuermann}, {Dreizler}, {Hessman}, \&
  {Deller}}]{Beuermann12}
{Beuermann}, K., {Dreizler}, S., {Hessman}, F.~V., \& {Deller}, J. 2012, \aap,
  543, A138

\bibitem[{{Beuermann} {et~al.}(2010){Beuermann}, {Hessman}, {Dreizler},
  {Marsh}, {Parsons}, {Winget}, {Miller}, {Schreiber}, {Kley}, {Dhillon},
  {Littlefair}, {Copperwheat}, \& {Hermes}}]{Beuermann10}
{Beuermann}, K. {et~al.} 2010, \aap, 521, L60

\bibitem[{{Blake} {et~al.}(2008){Blake}, {Bloom}, {Latham}, {Szentgyorgyi},
  {Skrutskie}, {Falco}, \& {Starr}}]{Blake08}
{Blake}, C.~H., {Bloom}, J.~S., {Latham}, D.~W., {Szentgyorgyi}, A.~H.,
  {Skrutskie}, M.~F., {Falco}, E.~E., \& {Starr}, D.~S. 2008, \pasp, 120, 860

\bibitem[{{Bolmont} {et~al.}(2011){Bolmont}, {Raymond}, \&
  {Leconte}}]{Bolmont11}
{Bolmont}, E., {Raymond}, S.~N., \& {Leconte}, J. 2011, ArXiv e-prints

\bibitem[{{Bond} {et~al.}(2010){Bond}, {O'Brien}, \& {Lauretta}}]{Bond10}
{Bond}, J.~C., {O'Brien}, D.~P., \& {Lauretta}, D.~S. 2010, \apj, 715, 1050

\bibitem[{{Boshkayev} {et~al.}(2012){Boshkayev}, {Rueda}, {Ruffini}, \&
  {Siutsou}}]{Boshkayev12}
{Boshkayev}, K., {Rueda}, J.~A., {Ruffini}, R., \& {Siutsou}, I. 2012, ArXiv
  e-prints

\bibitem[{{Burrows} {et~al.}(1997){Burrows}, {Marley}, {Hubbard}, {Lunine},
  {Guillot}, {Saumon}, {Freedman}, {Sudarsky}, \& {Sharp}}]{Burrows97}
{Burrows}, A. {et~al.} 1997, Astrophys.~J., 491, 856

\bibitem[{{Burrows} {et~al.}(2006){Burrows}, {Sudarsky}, \&
  {Hubeny}}]{Burrows06}
{Burrows}, A., {Sudarsky}, D., \& {Hubeny}, I. 2006, \apj, 640, 1063

\bibitem[{{Charpinet} {et~al.}(2011){Charpinet}, {Fontaine}, {Brassard},
  {Green}, {van Grootel}, {Randall}, {Silvotti}, {Baran}, {{\O}stensen},
  {Kawaler}, \& {Telting}}]{Charpinet11}
{Charpinet}, S. {et~al.} 2011, Nature, 480, 496

\bibitem[{{Chassefiere}(1996)}]{Chassefiere96}
{Chassefiere}, E. 1996, \icarus, 124, 537

\bibitem[{{Chauvin} {et~al.}(2005){Chauvin}, {Lagrange}, {Dumas}, {Zuckerman},
  {Mouillet}, {Song}, {Beuzit}, \& {Lowrance}}]{Chauvin05}
{Chauvin}, G., {Lagrange}, A.-M., {Dumas}, C., {Zuckerman}, B., {Mouillet}, D.,
  {Song}, I., {Beuzit}, J.-L., \& {Lowrance}, P. 2005, \aap, 438, L25

\bibitem[{{Di Stefano} {et~al.}(2010){Di Stefano}, {Howell}, \&
  {Kawaler}}]{DiStefano10}
{Di Stefano}, R., {Howell}, S.~B., \& {Kawaler}, S.~D. 2010, \apj, 712, 142

\bibitem[{{Domagal-Goldman} {et~al.}(2012){Domagal-Goldman}, {Segura},
  {Claire}, \& {Meadows}}]{DomagalGoldman12}
{Domagal-Goldman}, S.~D., {Segura}, A., {Claire}, M.~C., \& {Meadows}, V.~S.
  2012, Science, submitted

\bibitem[{{Dressing} {et~al.}(2010){Dressing}, {Spiegel}, {Scharf}, {Menou}, \&
  {Raymond}}]{Dressing10}
{Dressing}, C.~D., {Spiegel}, D.~S., {Scharf}, C.~A., {Menou}, K., \&
  {Raymond}, S.~N. 2010, \apj, 721, 1295

\bibitem[{{Erkaev} {et~al.}(2007){Erkaev}, {Kulikov}, {Lammer}, {Selsis},
  {Langmayr}, {Jaritz}, \& {Biernat}}]{Erkaev07}
{Erkaev}, N.~V., {Kulikov}, Y.~N., {Lammer}, H., {Selsis}, F., {Langmayr}, D.,
  {Jaritz}, G.~F., \& {Biernat}, H.~K. 2007, Astro.~\& Astrophys., 472, 329

\bibitem[{{Faedi} {et~al.}(2009){Faedi}, {West}, {Burleigh}, {Goad}, \&
  {Hebb}}]{Faedi09}
{Faedi}, F., {West}, R., {Burleigh}, M.~R., {Goad}, M.~R., \& {Hebb}, L. 2009,
  Journal of Physics Conference Series, 172, 012057

\bibitem[{{Farihi} {et~al.}(2012){Farihi}, {Subasavage}, {Nelan}, {Harris},
  {Dahn}, {Nordhaus}, \& {Spiegel}}]{Farihi12}
{Farihi}, J., {Subasavage}, J.~P., {Nelan}, E.~P., {Harris}, H.~C., {Dahn},
  C.~C., {Nordhaus}, J., \& {Spiegel}, D.~S. 2012, \mnras, 424, 519

\bibitem[{{Ferraz-Mello} {et~al.}(2008){Ferraz-Mello}, {Rodr{\'{\i}}guez}, \&
  {Hussmann}}]{FerrazMello08}
{Ferraz-Mello}, S., {Rodr{\'{\i}}guez}, A., \& {Hussmann}, H. 2008, Celestial
  Mechanics and Dynamical Astronomy, 101, 171

\bibitem[{{Finley} {et~al.}(1997){Finley}, {Koester}, \& {Basri}}]{Finley97}
{Finley}, D.~S., {Koester}, D., \& {Basri}, G. 1997, \apj, 488, 375

\bibitem[{{Fleming} {et~al.}(2010){Fleming}, {Ge}, {Mahadevan}, {Lee},
  {Eastman}, {Siverd}, {Gaudi}, {Niedzielski}, {Sivarani}, {Stassun},
  {Wolszczan}, {Barnes}, {Gary}, {Cuong Nguyen}, {Morehead}, {Wan}, {Zhao},
  {Liu}, {Guo}, {Kane}, {van Eyken}, {De Lee}, {Crepp}, {Shelden}, {Laws},
  {Wisniewski}, {Schneider}, {Pepper}, {Snedden}, {Pan}, {Bizyaev},
  {Brewington}, {Malanushenko}, {Malanushenko}, {Oravetz}, {Simmons}, \&
  {Watters}}]{Fleming10}
{Fleming}, S.~W. {et~al.} 2010, Astrophys.~J., 718, 1186

\bibitem[{{Fossati} {et~al.}(2012){Fossati}, {Bagnulo}, {Haswell}, {Patel},
  {Busuttil}, {Kowalski}, {Shulyak}, \& {Sterzik}}]{Fossati12}
{Fossati}, L., {Bagnulo}, S., {Haswell}, C.~A., {Patel}, M.~R., {Busuttil}, R.,
  {Kowalski}, P.~M., {Shulyak}, D.~V., \& {Sterzik}, M.~F. 2012, ArXiv e-prints

\bibitem[{{G{\"a}nsicke} {et~al.}(2006){G{\"a}nsicke}, {Marsh}, {Southworth},
  \& {Rebassa-Mansergas}}]{Gansicke06}
{G{\"a}nsicke}, B.~T., {Marsh}, T.~R., {Southworth}, J., \&
  {Rebassa-Mansergas}, A. 2006, Science, 314, 1908

\bibitem[{{Gizis} {et~al.}(2007){Gizis}, {Jao}, {Subasavage}, \&
  {Henry}}]{Gizis07}
{Gizis}, J.~E., {Jao}, W.-C., {Subasavage}, J.~P., \& {Henry}, T.~J. 2007,
  \apjl, 669, L45

\bibitem[{{Greenberg}(2009)}]{Greenberg09}
{Greenberg}, R. 2009, Astrophys.~J., 698, L42

\bibitem[{{Heller} {et~al.}(2010){Heller}, {Jackson}, {Barnes}, {Greenberg}, \&
  {Homeier}}]{Heller10}
{Heller}, R., {Jackson}, B., {Barnes}, R., {Greenberg}, R., \& {Homeier}, D.
  2010, Astro.~\& Astrophys., 514, A22+

\bibitem[{{Heller} {et~al.}(2011){Heller}, {Leconte}, \& {Barnes}}]{Heller11}
{Heller}, R., {Leconte}, J., \& {Barnes}, R. 2011, Astro.~\& Astrophys., 528,
  A27+

\bibitem[{{Hernanz} {et~al.}(1994){Hernanz}, {Garcia-Berro}, {Isern},
  {Mochkovitch}, {Segretain}, \& {Chabrier}}]{Hernanz94}
{Hernanz}, M., {Garcia-Berro}, E., {Isern}, J., {Mochkovitch}, R., {Segretain},
  L., \& {Chabrier}, G. 1994, Astrophys.~J., 434, 652

\bibitem[{{Horner} {et~al.}(2012){Horner}, {Wittenmyer}, {Hinse}, \&
  {Tinney}}]{Horner12}
{Horner}, J., {Wittenmyer}, R.~A., {Hinse}, T.~C., \& {Tinney}, C.~G. 2012,
  \mnras, 425, 749

\bibitem[{{Jackson} {et~al.}(2008){Jackson}, {Barnes}, \&
  {Greenberg}}]{Jackson08_hab}
{Jackson}, B., {Barnes}, R., \& {Greenberg}, R. 2008,
  Mon.~Not.~Roy.~Astron.~Soc., 391, 237

\bibitem[{{Jackson} {et~al.}(2009){Jackson}, {Barnes}, \&
  {Greenberg}}]{Jackson09}
---. 2009, Astrophys.~J., 698, 1357

\bibitem[{{Jayawardhana} {et~al.}(2003){Jayawardhana}, {Ardila}, {Stelzer}, \&
  {Haisch}}]{Jayawardhana03}
{Jayawardhana}, R., {Ardila}, D.~R., {Stelzer}, B., \& {Haisch}, Jr., K.~E.
  2003, Astron.~J., 126, 1515

\bibitem[{{Jura}(2003)}]{Jura03}
{Jura}, M. 2003, Astrophys.~J., 584, L91

\bibitem[{{Jura} \& {Xu}(2012)}]{JuraXu12}
{Jura}, M., \& {Xu}, S. 2012, Astron.~J., 143, 6

\bibitem[{{Kasting} {et~al.}(1993){Kasting}, {Whitmire}, \&
  {Reynolds}}]{Kasting93}
{Kasting}, J.~F., {Whitmire}, D.~P., \& {Reynolds}, R.~T. 1993, Icarus, 101,
  108

\bibitem[{{Kawaler}(2004)}]{Kawaler04}
{Kawaler}, S.~D. 2004, in IAU Symposium, Vol. 215, Stellar Rotation, ed.
  A.~{Maeder} \& P.~{Eenens}, 561

\bibitem[{{Kirkpatrick} {et~al.}(2011){Kirkpatrick}, {Cushing}, {Gelino},
  {Griffith}, {Skrutskie}, {Marsh}, {Wright}, {Mainzer}, {Eisenhardt},
  {McLean}, {Thompson}, {Bauer}, {Benford}, {Bridge}, {Lake}, {Petty},
  {Stanford}, {Tsai}, {Bailey}, {Beichman}, {Bloom}, {Bochanski}, {Burgasser},
  {Capak}, {Cruz}, {Hinz}, {Kartaltepe}, {Knox}, {Manohar}, {Masters},
  {Morales-Calder{\'o}n}, {Prato}, {Rodigas}, {Salvato}, {Schurr}, {Scoville},
  {Simcoe}, {Stapelfeldt}, {Stern}, {Stock}, \& {Vacca}}]{Kirkpatrick11}
{Kirkpatrick}, J.~D. {et~al.} 2011, \apjs, 197, 19

\bibitem[{{Koester}(2010)}]{Koester10}
{Koester}, D. 2010, \memsai, 81, 921

\bibitem[{{Koriski} \& {Zucker}(2011)}]{KoriskiZucker11}
{Koriski}, S., \& {Zucker}, S. 2011, Astrophys.~J., 741, L23

\bibitem[{{Lambeck}(1977)}]{Lambeck77}
{Lambeck}, K. 1977, Royal Society of London Philosophical Transactions Series
  A, 287, 545

\bibitem[{{Lammer} {et~al.}(2009){Lammer}, {Odert}, {Leitzinger},
  {Khodachenko}, {Panchenko}, {Kulikov}, {Zhang}, {Lichtenegger}, {Erkaev},
  {Wuchterl}, {Micela}, {Penz}, {Biernat}, {Weingrill}, {Steller}, {Ottacher},
  {Hasiba}, \& {Hanslmeier}}]{Lammer09}
{Lammer}, H. {et~al.} 2009, \aap, 506, 399

\bibitem[{{Leconte} {et~al.}(2010){Leconte}, {Chabrier}, {Baraffe}, \&
  {Levrard}}]{Leconte10}
{Leconte}, J., {Chabrier}, G., {Baraffe}, I., \& {Levrard}, B. 2010, Astro.~\&
  Astrophys., 516, A64+

\bibitem[{{Lee} {et~al.}(2011){Lee}, {Ge}, {Fleming}, {Stassun}, {Gaudi},
  {Barnes}, {Mahadevan}, {Eastman}, {Wright}, {Siverd}, {Gary}, {Ghezzi},
  {Laws}, {Wisniewski}, {Porto de Mello}, {Ogando}, {Maia}, {Nicolaci da
  Costa}, {Sivarani}, {Pepper}, {Cuong Nguyen}, {Hebb}, {De Lee}, {Wang},
  {Wan}, {Zhao}, {Chang}, {Groot}, {Varosi}, {Hearty}, {Hanna}, {van Eyken},
  {Kane}, {Agol}, {Bizyaev}, {Bochanski}, {Brewington}, {Chen}, {Costello},
  {Dou}, {Eisenstein}, {Fletcher}, {Ford}, {Guo}, {Holtzman}, {Jiang}, {French
  Leger}, {Liu}, {Long}, {Malanushenko}, {Malanushenko}, {Malik}, {Oravetz},
  {Pan}, {Rohan}, {Schneider}, {Shelden}, {Snedden}, {Simmons}, {Weaver},
  {Weinberg}, \& {Xie}}]{Lee11}
{Lee}, B.~L. {et~al.} 2011, Astrophys.~J., 728, 32

\bibitem[{{L{\'e}ger} {et~al.}(2004){L{\'e}ger}, {Selsis}, {Sotin}, {Guillot},
  {Despois}, {Mawet}, {Ollivier}, {Lab{\`e}que}, {Valette}, {Brachet},
  {Chazelas}, \& {Lammer}}]{Leger04}
{L{\'e}ger}, A. {et~al.} 2004, Icarus, 169, 499

\bibitem[{{Mainzer} {et~al.}(2011){Mainzer}, {Cushing}, {Skrutskie}, {Gelino},
  {Kirkpatrick}, {Jarrett}, {Masci}, {Marley}, {Saumon}, {Wright}, {Beaton},
  {Dietrich}, {Eisenhardt}, {Garnavich}, {Kuhn}, {Leisawitz}, {Marsh},
  {McLean}, {Padgett}, \& {Rueff}}]{Mainzer11}
{Mainzer}, A. {et~al.} 2011, Astrophys.~J., 726, 30

\bibitem[{{Metcalfe} {et~al.}(2004){Metcalfe}, {Montgomery}, \&
  {Kanaan}}]{Metcalfe04}
{Metcalfe}, T.~S., {Montgomery}, M.~H., \& {Kanaan}, A. 2004, Astrophys.~J.,
  605, L133

\bibitem[{{Monteiro}(2010)}]{Monteiro10}
{Monteiro}, H. 2010, Bulletin of the Astronomical Society of Brazil, 29, 22

\bibitem[{{Morbidelli} {et~al.}(2000){Morbidelli}, {Chambers}, {Lunine},
  {Petit}, {Robert}, {Valsecchi}, \& {Cyr}}]{Morbidelli00}
{Morbidelli}, A., {Chambers}, J., {Lunine}, J.~I., {Petit}, J.~M., {Robert},
  F., {Valsecchi}, G.~B., \& {Cyr}, K.~E. 2000, Meteoritics and Planetary
  Science, 35, 1309

\bibitem[{{Mullally} {et~al.}(2009){Mullally}, {Reach}, {De Gennaro}, \&
  {Burrows}}]{Mullally09}
{Mullally}, F., {Reach}, W.~T., {De Gennaro}, S., \& {Burrows}, A. 2009, \apj,
  694, 327

\bibitem[{{Mullally} {et~al.}(2008){Mullally}, {Winget}, {De Gennaro},
  {Jeffery}, {Thompson}, {Chandler}, \& {Kepler}}]{Mullally08}
{Mullally}, F., {Winget}, D.~E., {De Gennaro}, S., {Jeffery}, E., {Thompson},
  S.~E., {Chandler}, D., \& {Kepler}, S.~O. 2008, \apj, 676, 573

\bibitem[{{Nagasawa} {et~al.}(2008){Nagasawa}, {Ida}, \& {Bessho}}]{Nagasawa08}
{Nagasawa}, M., {Ida}, S., \& {Bessho}, T. 2008, \apj, 678, 498

\bibitem[{{Nordhaus} {et~al.}(2010){Nordhaus}, {Spiegel}, {Ibgui}, {Goodman},
  \& {Burrows}}]{Nordhaus10}
{Nordhaus}, J., {Spiegel}, D.~S., {Ibgui}, L., {Goodman}, J., \& {Burrows}, A.
  2010, \mnras, 408, 631

\bibitem[{{Pierrehumbert}(2010)}]{Pierrehumbert10}
{Pierrehumbert}, R.~T. 2010, {Principles of Planetary Climate}

\bibitem[{{Piro}(2011)}]{Piro11}
{Piro}, A.~L. 2011, \apjl, 740, L53

\bibitem[{{Pizzolato} {et~al.}(2003){Pizzolato}, {Maggio}, {Micela},
  {Sciortino}, \& {Ventura}}]{Pizzolato03}
{Pizzolato}, N., {Maggio}, A., {Micela}, G., {Sciortino}, S., \& {Ventura}, P.
  2003, \aap, 397, 147

\bibitem[{{Pollack} {et~al.}(1993){Pollack}, {Hurter}, \&
  {Johnson}}]{Pollack93}
{Pollack}, H.~N., {Hurter}, S.~J., \& {Johnson}, J.~R. 1993, Reviews of
  Geophysics, 31, 267

\bibitem[{{Prodan} \& {Murray}(2012)}]{ProdanMurray12}
{Prodan}, S., \& {Murray}, N. 2012, \apj, 747, 4

\bibitem[{{Raymond} {et~al.}(2004){Raymond}, {Quinn}, \& {Lunine}}]{Raymond04}
{Raymond}, S.~N., {Quinn}, T., \& {Lunine}, J.~I. 2004, Icarus, 168, 1

\bibitem[{{Renedo} {et~al.}(2010){Renedo}, {Althaus}, {Miller Bertolami},
  {Romero}, {C{\'o}rsico}, {Rohrmann}, \& {Garc{\'{\i}}a-Berro}}]{Renedo10}
{Renedo}, I., {Althaus}, L.~G., {Miller Bertolami}, M.~M., {Romero}, A.~D.,
  {C{\'o}rsico}, A.~H., {Rohrmann}, R.~D., \& {Garc{\'{\i}}a-Berro}, E. 2010,
  \apj, 717, 183

\bibitem[{{Ribas} {et~al.}(2005){Ribas}, {Guinan}, {G{\"u}del}, \&
  {Audard}}]{Ribas05}
{Ribas}, I., {Guinan}, E.~F., {G{\"u}del}, M., \& {Audard}, M. 2005, \apj, 622,
  680

\bibitem[{{Salpeter}(1961)}]{Salpeter61}
{Salpeter}, E.~E. 1961, Astrophys.~J., 134, 669

\bibitem[{{Segretain} {et~al.}(1994){Segretain}, {Chabrier}, {Hernanz},
  {Garcia-Berro}, {Isern}, \& {Mochkovitch}}]{Segratain94}
{Segretain}, L., {Chabrier}, G., {Hernanz}, M., {Garcia-Berro}, E., {Isern},
  J., \& {Mochkovitch}, R. 1994, Astrophys.~J., 434, 641

\bibitem[{{Selsis} {et~al.}(2007){Selsis}, {Kasting}, {Levrard}, {Paillet},
  {Ribas}, \& {Delfosse}}]{Selsis07}
{Selsis}, F., {Kasting}, J.~F., {Levrard}, B., {Paillet}, J., {Ribas}, I., \&
  {Delfosse}, X. 2007, Astro.~\& Astrophys., 476, 1373

\bibitem[{{Skemer} {et~al.}(2011){Skemer}, {Close}, {Sz{\H u}cs}, {Apai},
  {Pascucci}, \& {Biller}}]{Skemer11}
{Skemer}, A.~J., {Close}, L.~M., {Sz{\H u}cs}, L., {Apai}, D., {Pascucci}, I.,
  \& {Biller}, B.~A. 2011, \apj, 732, 107

\bibitem[{{Song} {et~al.}(2006){Song}, {Schneider}, {Zuckerman}, {Farihi},
  {Becklin}, {Bessell}, {Lowrance}, \& {Macintosh}}]{Song06}
{Song}, I., {Schneider}, G., {Zuckerman}, B., {Farihi}, J., {Becklin}, E.~E.,
  {Bessell}, M.~S., {Lowrance}, P., \& {Macintosh}, B.~A. 2006, \apj, 652, 724

\bibitem[{{Sotin} {et~al.}(2007){Sotin}, {Grasset}, \& {Mocquet}}]{Sotin07}
{Sotin}, C., {Grasset}, O., \& {Mocquet}, A. 2007, Icarus, 191, 337

\bibitem[{{Spiegel} {et~al.}(2011){Spiegel}, {Burrows}, \&
  {Milsom}}]{Spiegel11}
{Spiegel}, D.~S., {Burrows}, A., \& {Milsom}, J.~A. 2011, \apj, 727, 57

\bibitem[{{Stassun} {et~al.}(2006){Stassun}, {Mathieu}, \&
  {Valenti}}]{Stassun06}
{Stassun}, K.~G., {Mathieu}, R.~D., \& {Valenti}, J.~A. 2006, Nature, 440, 311

\bibitem[{{Strom} {et~al.}(1979){Strom}, {Terrile}, {Hansen}, \&
  {Masursky}}]{Strom79}
{Strom}, R.~G., {Terrile}, R.~J., {Hansen}, C., \& {Masursky}, H. 1979, Nature,
  280, 733

\bibitem[{{Veeder} {et~al.}(1994){Veeder}, {Matson}, {Johnson}, {Blaney}, \&
  {Goguen}}]{Veeder94}
{Veeder}, G.~J., {Matson}, D.~L., {Johnson}, T.~V., {Blaney}, D.~L., \&
  {Goguen}, J.~D. 1994, J.~Geophys.~Res., 991, 17095

\bibitem[{{Watson} {et~al.}(1981){Watson}, {Donahue}, \& {Walker}}]{Watson81}
{Watson}, A.~J., {Donahue}, T.~M., \& {Walker}, J.~C.~G. 1981, Icarus, 48, 150

\bibitem[{{Williams} {et~al.}(1997){Williams}, {Kasting}, \&
  {Wade}}]{Williams97}
{Williams}, D.~M., {Kasting}, J.~F., \& {Wade}, R.~A. 1997, Nature, 385, 234

\bibitem[{{Williams} \& {Pollard}(2002)}]{WilliamsPollard02}
{Williams}, D.~M., \& {Pollard}, D. 2002, International Journal of
  Astrobiology, 1, 61

\bibitem[{{Yelle}(2004)}]{Yelle04}
{Yelle}, R.~V. 2004, \icarus, 170, 167

\bibitem[{{Yoder}(1995)}]{Yoder95}
{Yoder}, C.~F. 1995, in Global Earth Physics: A Handbook of Physical Constants,
  ed. {T.~J.~Ahrens}, 1--+

\end{thebibliography}

\end{document}